\documentclass[aip,jcp,amsmath,amssymb,reprint]{revtex4-1}
\usepackage[dvipdfmx]{graphicx}
\usepackage{bmpsize}
\usepackage{epsf}
\usepackage{bm}
\usepackage[usenames,dvipsnames]{xcolor}
\usepackage[normalem]{ulem}

\def\ls{\lesssim}

\begin{document}

\title{
How do hydrogen bonds break in supercooled water?: Detecting pathways not going through
saddle point of two-dimensional potential of mean force
}

\author{Takuma Kikutsuji}
\affiliation{
Division of Chemical Engineering,
Graduate School of Engineering Science, Osaka University, Toyonaka, Osaka 560-8531, Japan
}

\author{Kang Kim}
\email{kk@cheng.es.osaka-u.ac.jp}
\affiliation{
Division of Chemical Engineering,
Graduate School of Engineering Science, Osaka University, Toyonaka, Osaka 560-8531, Japan
}

\author{Nobuyuki Matubayasi}
\email{nobuyuki@cheng.es.osaka-u.ac.jp}
\affiliation{
Division of Chemical Engineering,
Graduate School of Engineering Science, Osaka University, Toyonaka, Osaka 560-8531, Japan
}
\affiliation{
Elements Strategy Initiative for Catalysts and Batteries, Kyoto
University, Katsura, Kyoto 615-8520, Japan
}

\date{\today}

\begin{abstract}
Supercooled water exhibits remarkably slow dynamics
similar to the behavior observed for various glass-forming liquids.
The local order of tetrahedral structures due to hydrogen-bonds (H-bonds)
 increases with decreasing temperature.
Thus, it is important to clarify the temperature dependence of the
H-bond breakage process.
This was investigated here using molecular dynamics simulations of 
TIP4P supercooled water.
The two-dimensional (2D) potential of mean force (PMF)
is presented
 using combinations of intermolecular distance and
 angle between two water molecules.
The saddle point of the 2D PMF suggests the presence of the transition state that
distinguishes between H-bond and non H-bond states.
However, we observed pathways not going through this saddle point
particularly at supercooled states, 
which are due to translational, rather than rotational motions of the molecules.
We quantified the characteristic time scales of rotational and
 translational H-bond breakages.
The time scale of translational H-bond breakage shows a non-Arrhenius
 temperature dependence comparable to that of the H-bond lifetime.
This time scale is
 relevant for the temperature dependence of the transmission
 coefficient based on the transition state theory.
The translational H-bond breakage is also related to cage-jumps observed in
 glass-forming liquids, which mostly involve spatially correlated molecules.
Our findings 
warrant further exploration of an appropriate free-energy
 surface or reaction 
 coordinates beyond the geometrical variables of the water dimer to describe
a possible saddle point related to collective jump motions.
\end{abstract}

\maketitle

\section{Introduction}

Liquid water is a complex material that exhibits many anomalous properties.~\cite{eisenberg2005structure}
When liquid water is supercooled below its melting temperature, 
such anomalies become remarkable.
The controversial concept of a liquid-liquid transition in deeply
supercooled states has attracted
much attention from researchers.~\cite{Debenedetti:2003gn, Stanley:2013gw,
Gallo:2016fd}
Thus, clarification of the structures and dynamics in supercooled water
has been increasingly important in recent years.
In particular, water molecules in supercooled water exhibit remarkably
slow dynamics similar to
that of viscous glass-forming liquids.~\cite{Gallo:1996hf,
Sciortino:1996hk, Sciortino:1997iw, Giovambattista:2003gz,
Giovambattista:2005ib, Chen:2006kk, Kumar:2007hl, Xu:2009hq, Gallo:2012cz, Dehaoui:2015ii,
DeMarzio:2016hl, Guillaud:2016bk, Galamba:2017eq, DeMarzio:2017fa, 
Kawasaki:2017gw, Singh:2017cc}

It is generally expected that the hydrogen-bonds (H-bonds) of water molecules
play a crucial role in determining 
their anomalous properties.~\cite{eisenberg2005structure}
A large number of experiments and simulations have been carried out to
investigate the local structure of H-bonds and the network
rearrangement.~\cite{Stillinger:1975fa, Stillinger:1980ws, Ohmine:1993ij,
Teixeira:2006eb, Bakker:2010ji, Agmon:2012jd, Perakis:2016gx}
The H-bond is generally defined based on 
certain structural or energetic criteria of the water-water configuration.
Various H-bond definitions in liquid water have been developed for molecular
dynamics (MD) simulations.~\cite{Kumar:2007bs, Matsumoto:2007cf,
PradaGracia:2013hc, Ozkanlar:2014gm}
A widely used criterion for determining the H-bond is a geometry definition
for a pair of water molecules, \textit{i.e.}, a pair of water molecules
is considered H-bonded if the intermolecular distance and angle become
less than pre-assigned threshold values.
Accordingly, the H-bond correlation function is formulated, allowing
quantification of the averaged H-bond lifetime,
$\tau_\mathrm{HB}$.~\cite{Stillinger:1975fa, Rapaport:1983ib,
Luzar:1996gx, Luzar:1996gw, Luzar:2000gv}
Its derivative with respect to time, which is related to the
reactive flux, characterizes the H-bond breakage
rate.~\cite{Luzar:1996gx, Luzar:1996gw, Luzar:2000gv}
As an alternative to $\tau_\mathrm{HB}$ from the correlation function,
the distribution function of the H-bond lifetime has been examined
using the trajectory based analyses.~\cite{Sciortino:1990el,
Starr:1999ki, Starr:2000cz, Henchman:2010ky, Martiniano:2013ck, Galamba:2017eq}
Furthermore, the molecular mechanism of the breakage and reforming of H-bond was
comprehensively investigated considering molecular mobility of bifurcated
H-bonds~\cite{Sciortino:1991hh, Sciortino:1992eu} and
molecular reorientational motions.~\cite{Laage:2006jw,
Laage:2008he, Henchman:2010ky, Stirnemann:2012co, Henchman:2016bm}.

The analysis of H-bond dynamics was also applied in MD simulations of supercooled
water.~\cite{Starr:1999ki, Starr:2000cz, Stirnemann:2012co, Martiniano:2013ck,
Saito:2013ff, Kawasaki:2017gw}
The temperature dependence is conventionally analyzed by examining the Arrhenius plot,
$\tau_\mathrm{HB}\propto \exp(E_\mathrm{A}/k_\mathrm{B}T)$, where $T$ is
the temperature, $k_\mathrm{B}$ is the Boltzmann constant, and
$E_\mathrm{A}$ is the
Arrhenius activation energy of H-bond breakage considering reaction rate theory.
The MD studies revealed that $\tau_\mathrm{HB}$ significantly
increased with decreasing temperature,
showing non-Arrhenius behavior, where $E_\mathrm{A}$ increased with decreasing
the temperature.
Collective molecular motions are expected to be a possible scenario
by using the analogy with dynamic heterogeneities in glassy systems.~\cite{Berthier:2011wv}
However, generally speaking, it is difficult to interpret
for $E_\mathrm{A}$ of supercooled liquids and glasses,
\textit{i.e.}, a fragility classification for the temperature dependence of the
dynamics.~\cite{Angell:2008hg}
The fragility is also relevant with searching for transition states connecting
numerous stable states in the rugged free-energy landscape of complex
many-body systems.~\cite{Stillinger:1995fu, Debenedetti:2001bh}
A study based on this concept was reported using
configuration-space-network analysis for H-bond rearrangements.~\cite{PradaGracia:2012hp}

Kumar \textit {et al.} proposed a method to describe the profile for the two-dimensional (2D)
potential of mean force (PMF), which is also referred to as the
free-energy surface, from the distribution function of the 
intermolecular distance and angle between two water molecules.~\cite{Kumar:2007bs}
This 2D PMF profile enabled the systematic quantification of the 
distance and angle thresholds, which distinguishes between H-bond and
non H-bond regions in classical MD simulations.
In Ref.~\onlinecite{Kumar:2007bs}, an attempt was made to understand the consistency
with H-bond definition based on an electronic structure of water dimer.
In addition, profiles of 2D PMF have been also provided by
recent ab initio MD simulations.~\cite{Morawietz:2016cp, Chen:2017jn}
If the chosen intermolecular distance and angle are treated as
the reaction coordinates for the H-bond breakage transition, 
the dynamics of H-bond breakage is then predicted by the pathway
that goes through the saddle point on the 2D PMF profile.

In general, 
it is challenging to 
appropriately describe the transition state in various chemical
processes in many-body systems,
including the H-bond breakage in liquid water investigated
here.~\cite{Bolhuis:1998cj, Geissler:1999kj, Bolhuis:2000eo}
Using a stochastic transition path sampling method for rare events,
the kinetic pathway of the H-bond breakage in liquid water was investigated, but
little attention was paid to the connection with the 2D PMF.~\cite{Csajka:1998et}
The aim of this study is to address the pathway of H-bond breakage,
particularly in supercooled water.
In addition, we discuss the 
mechanism of the non-Arrhenius behavior of the temperature dependence of 
$\tau_\mathrm{HB}$.

This study analyses the impact of the 
the H-bond breakage dynamics in TIP4P
supercooled water using MD simulations.
The investigated temperatures ranged from 300 K to 190 K at a fixed volume.
Using the geometrical variables between water dimer, 
we calculated the distance-angle
distribution function and the associated 2D PMF.~\cite{Kumar:2007bs}
From the 2D profile, H-bond and non H-bond regions are distinguished
by specifying the saddle point.
The H-bond lifetime was quantified from the H-bond correlation function.
In addition, the transmission coefficient was calculated based on
transition state theory (TST) from 
reactive flux analysis.~\cite{Chandler:1978hh, Hanggi:1990dk} and 
its relationship with the free-energy barrier of the saddle point on the
2D PMF profile was examined.

The present work also focuses on the characteristic time
scales of rotational and translational H-bond
breakages, which were evaluated from the time dependent H-bond breakage populations.
To this aim, the populations in the neighboring regions of the H-bond
region were quantified on the 2D PMF profile.
The physical implication 
of the transmission coefficient was examined considering
the relationship between rotational and
translational H-bond breakages.

\section{Model and simulations}
\label{sec:model}

MD simulations were performed using the TIP4P water
model.~\cite{Jorgensen:1983fl}
The various properties of this model have been intensely examined so
far.
In particular, the comparison with other models including TIP4P/2005 was
carefully performed.~\cite{Vega:2011bb}
All the simulations in this work were performed with the GROMACS
package.~\cite{Hess:2008db, Abraham:2015gj}
The simulation system contained
$N=1,000$ molecules in the cubic box with the
periodic boundary conditions.
The mass density was fixed at $1$ $\mathrm{g/cm^{3}}$.
Correspondingly, the linear dimension of the system was approximately 3.1 nm.
The investigated temperatures were $T=300$, $260$, $240$, $220$, $210$,
$200$, and $190$ K.
The system was first equilibrated with the \textit{NVT} ensemble at each
temperature for 10 ns.
Then, the trajectories for the
calculations of various quantities were produced with the \textit{NVE}
ensemble for 10 ns ($T \ge 210$ K) and 100 ns ($T \le 200$ K).
A time step of 1 fs was used.
For this model, 
dynamical quantities such as intermediate scattering function
and mean square displacement have been reported
previously,~\cite{Gallo:2012cz} 
with which our simulation results were in agreement (data not shown).
This indicates that our computational setups are adequate in turn.

The H-bond was investigated by using distance-angle definitions between
two water molecules.~\cite{Kumar:2007bs}
Specifically, a pair of $R$ and $\beta$ was chosen, where
$R$ represents the O-O intermolecular distance and 
$\beta$ is the O-OH intermolecular angle.
Note that $0< \beta< 180^\circ$.
The combined distance-angle distribution function, $g(R, \beta)$, was
calculated at each temperature.~\cite{Kumar:2007bs}
For this $R$-$\beta$ definition, $2\pi\rho R^2 \sin\beta g(R, \beta)dRd\beta$ represents the
averaged number of O atoms found in
the partial spherical shell having $dR$ and $d\beta$ at distance $R$ and
angle $\beta$ from one fixed O atom.
Here, $\rho$ is the molecular density of the system.
The function $g(R, \beta)$ results in the PMF defined by $W(R,
\beta)=-k_\mathrm{B}T \ln g(R, \beta)$.
This 2D PMF can be regarded as the free-energy surface using
reaction coordinates $(R, \beta)$.
The saddle point of $W(R, \beta)$ was numerically determined from the
calculations of the gradient $\nabla W(R, \beta)$.
Furthermore, the free-energy difference $\Delta G^\ddagger$ between the
global minimum and saddle point was quantified.

We calculated the time correlation function of the H-bond, 
$
c(t) = \langle h(0) h(t)\rangle/ \langle h(0) \rangle,
$
where $h(t)$ denotes the H-bond operator at a time
$t$.~\cite{Luzar:1996gw, Luzar:1996gx}
Here, $\tau_\mathrm{HB}$ was determined from $c(t)$
by fitting it to
the exponential function
$\exp(-t/\tau_\mathrm{HB})$.\footnote{
Note that the stretched exponential function,
$\exp[-(t/\tau_\mathrm{HB})^{\beta_\mathrm{HB}}]$, provided a better
fitting for $c(t)$ with the exponent $\beta_\mathrm{HB}\approx 0.7$.
The obtained $\tau_\mathrm{HB}$ decreased by up to 15\% from the value
obtained with $\beta_\mathrm{HB}=1$.
However, the temperature dependence of $\tau_\mathrm{HB}$ was not
influenced overall.}
Furthermore, we examined the reactive flux function,
$
k(t) = -dc(t)/dt,
$
which quantifies the averaged rate of H-bond
breakage.~\cite{Luzar:1996gx, Luzar:2000gv}
In particular, $k(0)$ is the so-called TST rate constant,
$k_\mathrm{TST}$, which characterizes the escape rate towards the H-bond broken state
at the assumed transition state.
We estimated $k_\mathrm{TST}$ from a finite difference of $c(t)$ with
$\Delta t=1$ fs at each temperature.
In contrast, the rate constant for H-bond breakage is
approximated by $k= 1/\tau_\mathrm{HB}$, which corresponds to the
plateau value of $k(t)$ at longer times.
These two variables are connected by introducing the transmission
coefficient, defined as $\kappa = k /k_\mathrm{TST}$, where $\kappa$ is
generally less than unity.~\cite{Chandler:1978hh, Hanggi:1990dk}
The role of the transmission coefficient has been also discussed 
for H-bond kinetics in liquid water.~\cite{Luzar:2000gv}

\begin{figure}[t]
\centering
\includegraphics[width=0.48\textwidth]{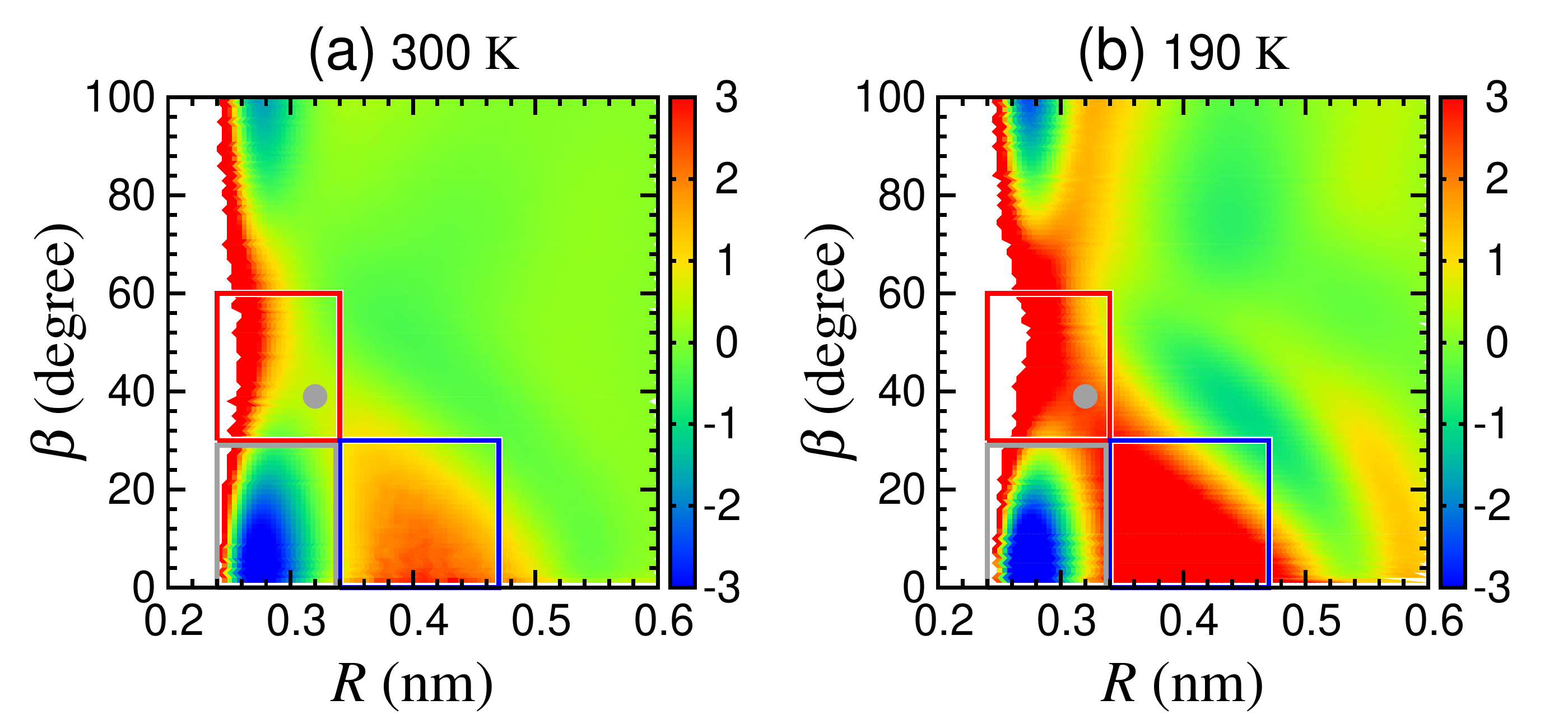}
\caption{
Contour plots of the potential of mean force, $W(R, \beta)$, at  a
 temperature of 300 K (a) and 190 K (b).
The value of the color bar is normalized by
 $k_\mathrm{B}T$.
The saddle point is shown by the gray point.
The rectangle indicated by the gray line represents the H-bond region.
The areas surrounded by red and blue lines represent the H-bond breakage
 regions due to rotational and translational motions, respectively
 (denoted as regions R and T).
The region with unsampled configurations is shown in white.
}
\label{fig:2D_PMF} 
\end{figure}

\section{Results and discussion}
\label{sec:results}

\subsection{2D PMF and H-bond lifetimes}
\label{sec:2D_PMF}

The 2D PMF contour plots are shown in Fig.~\ref{fig:2D_PMF} for
$T=300$ K and $190$ K.
The H-bond criterion is given by the
geometrical condition between the two water molecules.~\cite{Kumar:2007bs}
Two molecules are considered H-bonded
if the distance-angle relationship is 
$(0.24~\mathrm{nm}, 0^\circ) \le (R, \beta) \le (0.34~\mathrm{nm},
30^\circ)$ (rectangular area indicated in Fig.~\ref{fig:2D_PMF}).
This criterion is mostly consistent with that reported previously.~\cite{Kumar:2007bs}
At shorter distance less than $R=0.24$ nm, 
the radial distribution functions $g_\mathrm{OO}(r)$ vanishes.
The position $R=0.34$ nm corresponds to the first
minima of $g_\mathrm{OO}(r)$.
These threshold values remain unchanged at any temperature, as seen in Fig.~\ref{fig:2D_PMF}.
Figure~\ref{fig:2D_PMF} also shows that the temperature dependence of  
saddle point position is negligible.
Instead, the roughness of the landscape becomes larger with decreasing
the temperature and 
the free-energy difference $\Delta G^\ddagger$ between the
global minimum and saddle point becomes correspondingly larger.

The transition from H-bonded to H-bond breakage states can be generally characterized
by the saddle point on the free-energy surface.
From our calculations, $(R^\ddagger, \beta^\ddagger)=(0.32~\mathrm{nm}, 39^\circ)$
was obtained (indicated by the gray dot in Fig.~\ref{fig:2D_PMF}).
The H-bond is expected to be broken by increasing $\beta$,
 \textit{i.e.}, by exploiting molecular rotational motions.
It is important to note that these 2D plots are not directly linked
with $k_\mathrm{TST}$, which
is the reactive flux at $t=0$ for a barrier-crossing dynamics
with a dividing surface.
The introduction of $k_\mathrm{TST}$ corresponds
effectively to adopting a one-dimensional description of the reaction coordinate.
On the other hand, a focus in the present work is to determine the TST-type
expression, $\exp(-\Delta G^\ddagger/k_\mathrm{B}T)$, evaluated by
referring to the saddle point on the 2D PMF profile.
Although this expression does not definitely coincide with
$k_\mathrm{TST}$, 
$\Delta G^{\ddagger}$ can be regarded as the Arrhenius activation energy $E_\mathrm{A}$
when it is independent of temperature over the dividing
surface for the reactive flux and $\kappa$ is unity.
Accordingly, the deviation from the Arrhenius behavior of
$k=1/\tau_\mathrm{HB}$ has three sources: 
(1) the temperature dependence of $\Delta G^{\ddagger}$, 
which is concerned with the entropy of the activation; (2)
the dimensionality
of the reaction coordinate, which depends on the choice of reaction
coordinate; and (3)
the temperature dependence of $\kappa(< 1)$, which
reflects the detailed dynamics of barrier crossing.
These issues are discussed in more detail.

\begin{figure}[t]
\centering
\includegraphics[width=0.48\textwidth]{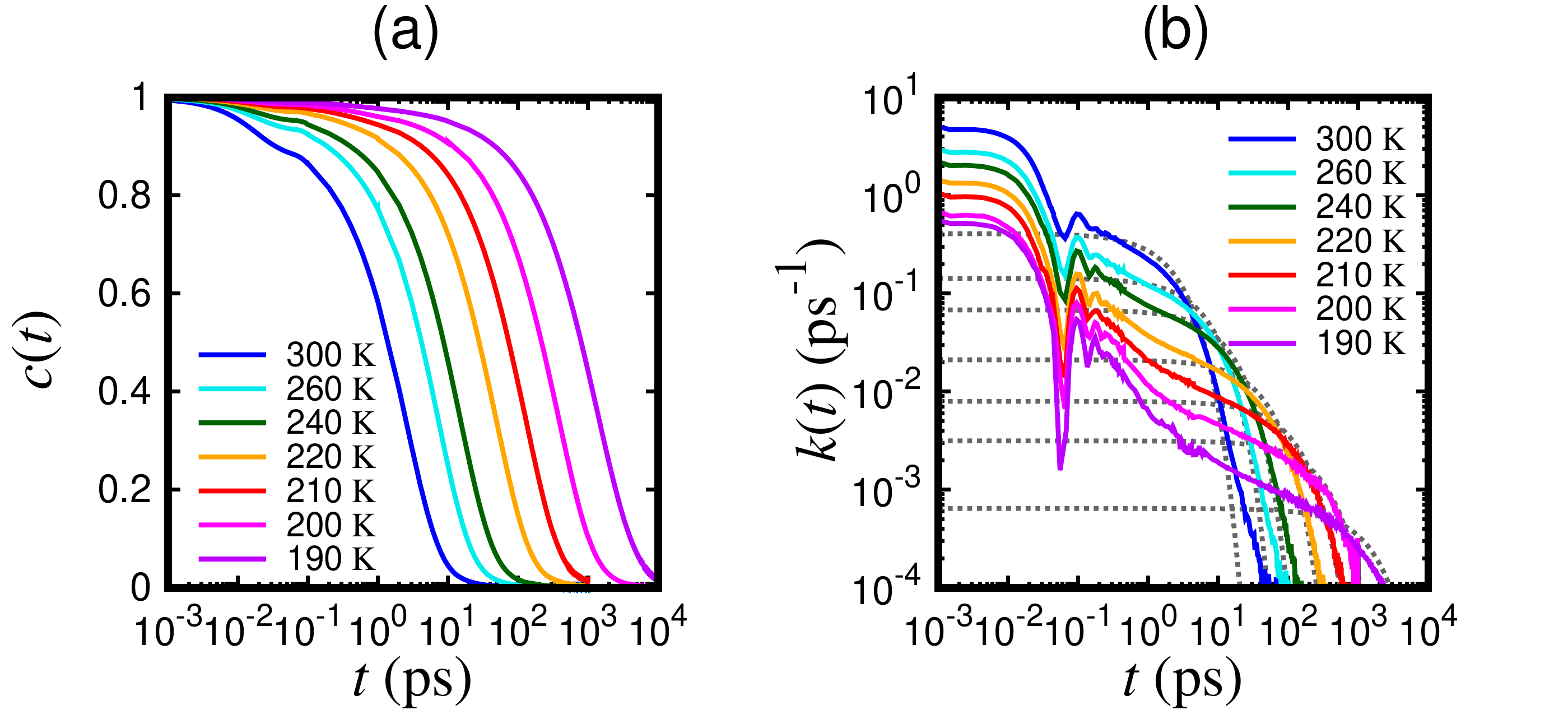}
\caption{
(a) H-bond correlation function $c(t)$.
(b) Reactive flux $k(t)=-dc(t)/dt$.
Dotted curves represent the exponential function
 $\exp(-t/\tau_\mathrm{HB})/\tau_\mathrm{HB}$ with H-bond lifetime $\tau_\mathrm{HB}$.}
\label{fig:correlation} 
\end{figure}

\begin{figure}[t]
\centering
\includegraphics[width=0.35\textwidth]{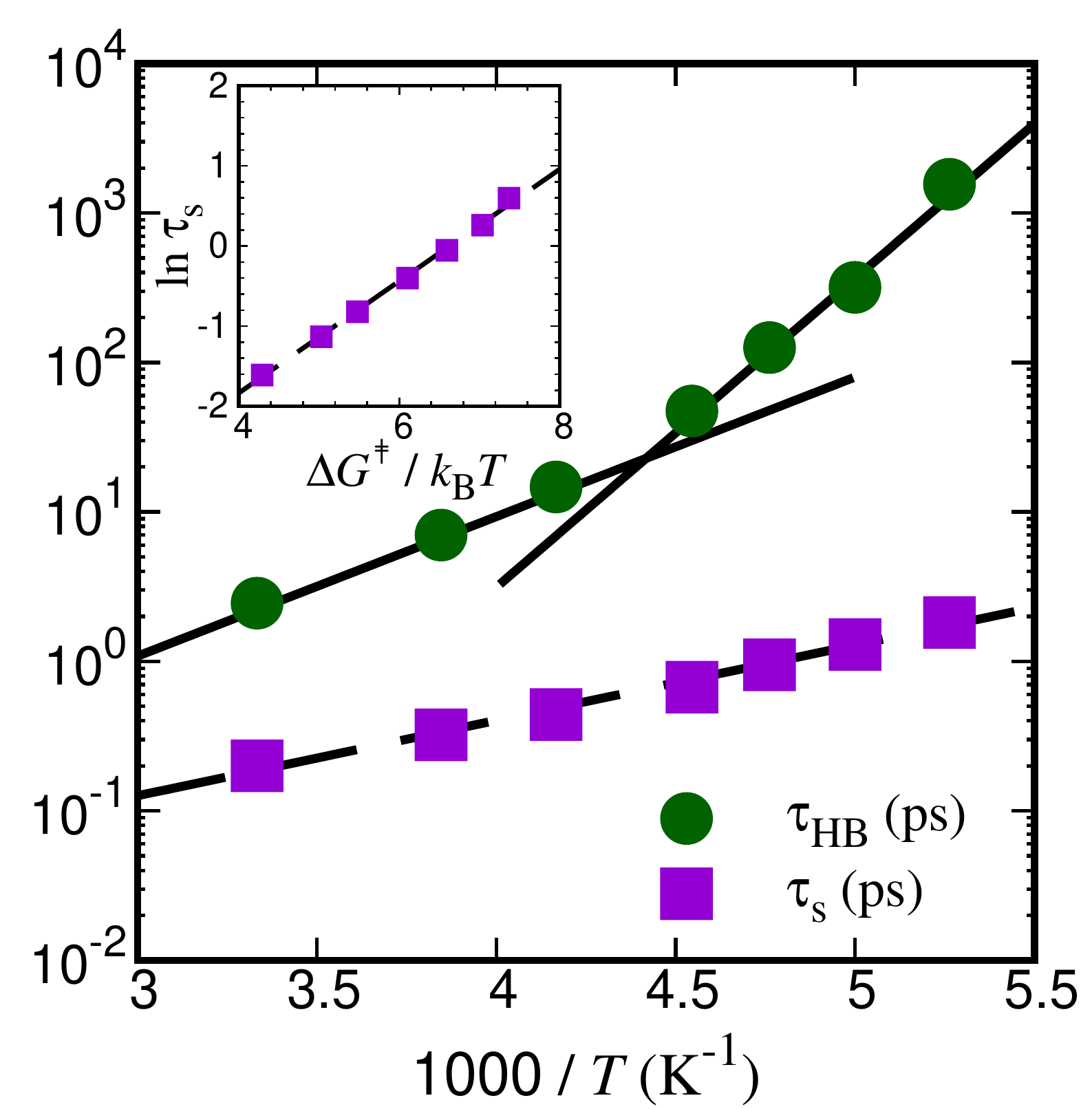}
\caption{
H-bond lifetime $\tau_\mathrm{HB}$ and inverse TST rate constant
 $\tau_s={k_\mathrm{TST}}^{-1}$ vs. inverse temperature $1000/T$.
The Arrhenius behavior $\tau_\mathrm{HB}\propto
 \exp(E_\mathrm{A}/k_\mathrm{B}T)$ in the 
 high and low temperature ranges are shown as two straight lines, with
 activation energies with $E_\mathrm{A} =$ 17.9 and 39.6 kJ/mol, respectively. 
The dashed line shows Arrhenius behavior for $\tau_\mathrm{s}\propto \exp(E_\mathrm{s}/k_\mathrm{B}T)$
with $E_\mathrm{s} =$ 9.7 kJ/mol.
Inset:
$\ln \tau_\mathrm{s}$
 vs. the free-energy barrier of the saddle point
 $\Delta G^\ddagger/k_\mathrm{B}T$ on the 2D PMF profile.
The dashed straight line is the fitting of $\tau_\mathrm{s}\propto \exp(C\Delta
 G^\ddagger/k_\mathrm{B}T)$ with a constant $C=0.7$.
}
\label{fig:tau_HB} 
\end{figure}

The H-bond correlation function $c(t)$ and reactive flux
$k(t)$ are shown in Fig.~\ref{fig:correlation}.
For comparison, the exponential decay curve
$\exp(-t/\tau_\mathrm{HB})/\tau_\mathrm{HB}$ is plotted in Fig.~\ref{fig:correlation}(b).
As mentioned in the end of Sec.~\ref{sec:model}, the decay of $k(t)$ at longer times is characterized by the
inverse of H-bond lifetime, $1/\tau_\mathrm{HB}$, at each temperature.
These behaviors are consistent with previously
reported results
using SPC/E supercooled water.~\cite{Starr:1999ki, Starr:2000cz}
The temperature dependence of 
$\tau_\mathrm{HB}$ is plotted in Fig.~\ref{fig:tau_HB}, where
$\tau_\mathrm{HB}$ increases significantly with decreasing temperature.
This drastic increase in $\tau_\mathrm{HB}$, particularly at supercooled
states, has been recently reported using MD simulations.~\cite{Kawasaki:2017gw}
In particular, the temperature dependence shows Arrhenius behaviors,
$\exp(E_\mathrm{A}/k_\mathrm{B}T)$, with different $E_\mathrm{A}$ values.
As shown in Fig.~\ref{fig:tau_HB}, 
$E_\mathrm{A}$ increases with decreasing the temperature.
The $E_\mathrm{A}$ were 
17.9 and 39.6 kJ/mol at the high and low temperature regions, respectively. 
In contrast, the time scale
$\tau_\mathrm{s}=1/k_\mathrm{TST}$ associated with the TST rate constant
shows the Arrhenius temperature dependence, $\tau_\mathrm{s}\propto \exp(E_\mathrm{s}/k_\mathrm{B}T)$.
The estimated Arrhenius activation energy was $E_\mathrm{s}=$ 9.7 kJ/mol, as shown in Fig.~\ref{fig:tau_HB}.
Similar values of the activation energy have reported in other liquid
water models, where the average of 
H-bond persistence lifetime was analyzed.~\cite{Starr:1999ki, Starr:2000cz,
Martiniano:2013ck, Galamba:2017eq}
This persistence lifetime is relevant with the time scale of TST rate constant.~\cite{Luzar:2000gv}
These numerical results are also comparable with experimental values
obtained from depolarized Rayleigh scattering.~\cite{Conde:1984dq,
Teixeira:1985dv}.
Our observations indicate that the transmission coefficient $\kappa=
k/k_\mathrm{TST}=\tau_\mathrm{s}/\tau_\mathrm{HB}$ decreases significantly
when the temperature is lowered.

Next, we investigated 
the relationship between $k_\mathrm{TST}$ and $\Delta
G^{\ddagger}$ (note that $k_\mathrm{TST}$ is
introduced as $k(0)$ in this work).
Inset of Fig.~\ref{fig:tau_HB} demonstrates
the TST-type relationship, $\tau_\mathrm{s}\propto \exp(C \Delta
G^\ddagger/k_\mathrm{B}T))$ with the slope about 0.7;
the free-energy barrier corresponding to $k_\mathrm{TST}$ is
underestimated considering the expected value of TST.
We again note that the free-energy barrier of 2D PMF is a different
quantity from that of the TST framework, which may have resulted in this
underestimation.
Another possible explanation is H-bond breakages due to
non-trivial environmental effects around the tagged H-bonded molecules.
Indeed, it has been demonstrated that 
molecular reorientations are occurred collectively.~\cite{Laage:2006jw}
Such collective motions may have decreased the activation barrier lower than
the TST prediction.

From the general expression, $k=\kappa k_\mathrm{TST}$,
the non-Arrhenius behavior of $\tau_\mathrm{HB}(= 1/k)$ and the
increase in $E_\mathrm{A}$ with decreasing 
temperature are attributed to the temperature dependency of 
transmission coefficient $\kappa$, which considerably decreases with
decreasing temperature (see Fig.~\ref{fig:tau_HB}).
An analogous non-Arrhenius temperature dependence of $\tau_\mathrm{HB}$
has been demonstrated in
TIP4P/2005 supercooled water.~\cite{Kawasaki:2017gw}
The physical implication of this is elucidated later.

\subsection{Temporal development of H-bond distribution function}
\label{sec:temporal}

\begin{figure*}[t]
\centering
\includegraphics[width=0.9\textwidth]{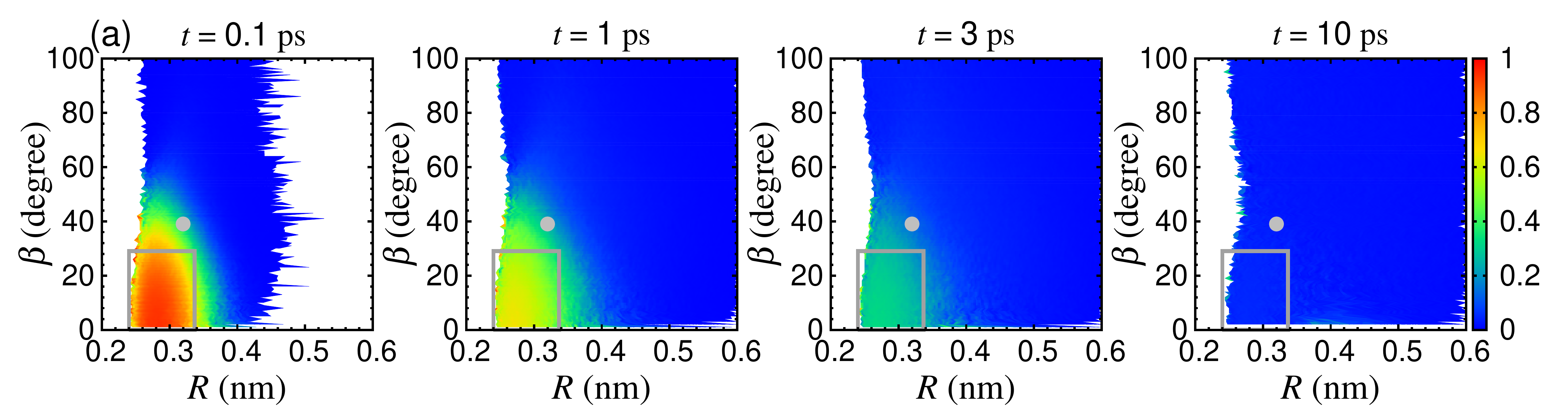}
\includegraphics[width=0.9\textwidth]{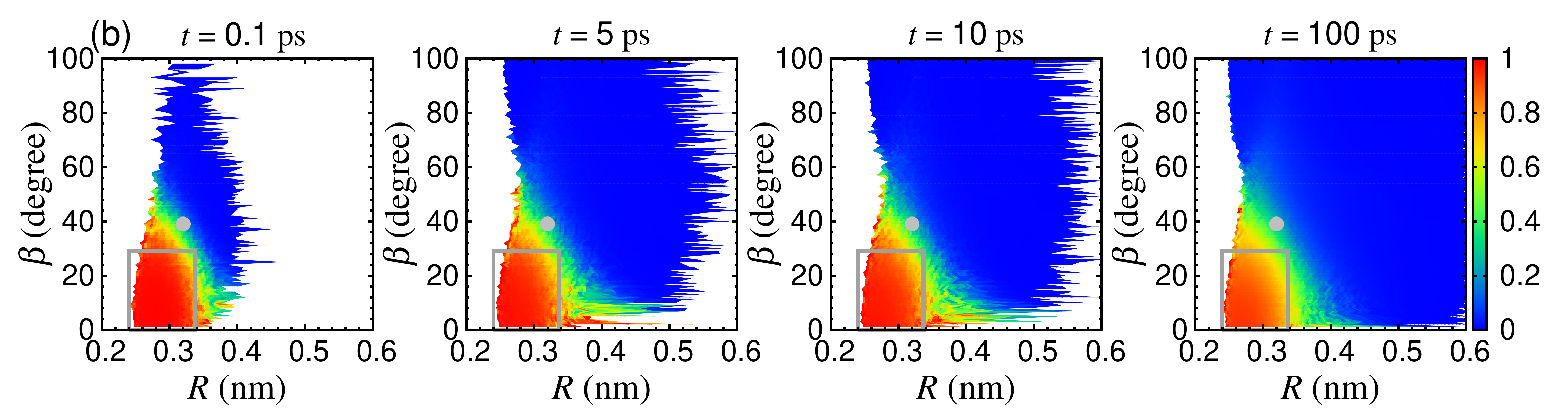}
\caption{Conditional distance-angle distribution function $G(R, \beta;
 t)$ at temperatures of 300 K (a) and 190 K (b).
The saddle point is shown by the gray point.
The rectangle indicated by the gray line represents the H-bond region.
The region with unsampled configurations is shown in white.
(Multimedia view)
} 
\label{fig:diff_oo_fig}
\end{figure*}

\begin{figure}[t]
\centering
\includegraphics[width=0.48\textwidth]{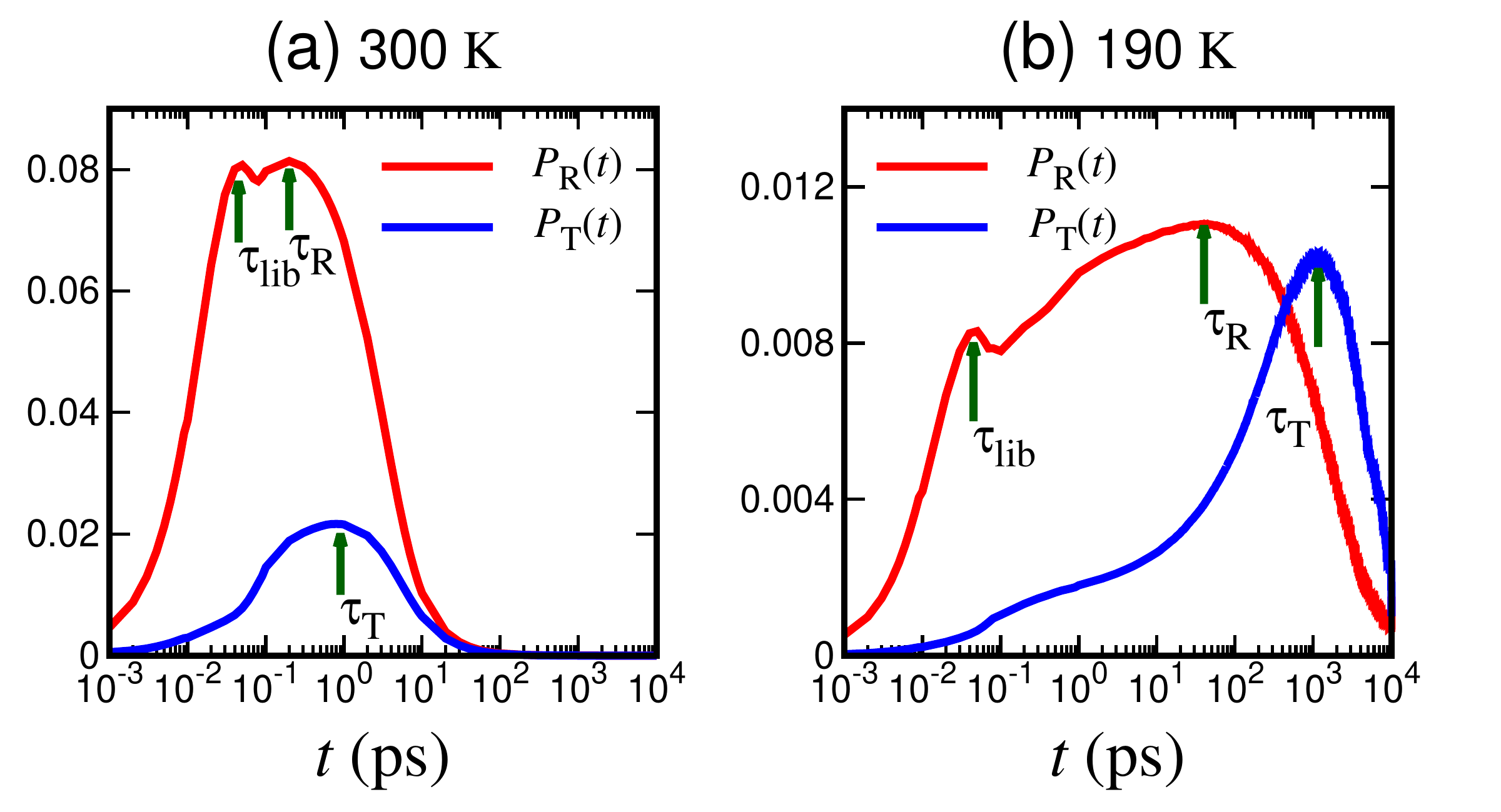}
\caption{
Time evolution of the normalized populations $P_\mathrm{R}(t)$
and $P_\mathrm{T}(t)$ for H-bond breakage regions at temperatures of 300 K (a)
 and 190 K (b).
The H-bond breakage regions (regions R and T) are depicted in Fig.~\ref{fig:2D_PMF}.
Here, $\tau_\mathrm{lib}$, $\tau_\mathrm{R}$, and $\tau_\mathrm{T}$ are indicated by arrows.
}
\label{fig:population}
\end{figure}

To reveal the molecular mechanism of H-bond breakage in supercooled water,
we examined the change of the geometric structure of two water molecules that are initially H-bonded.
To this end, we propose the extension of $g(R, \beta)$ to the time dependent
distribution function.
Specifically, we introduce the conditional 
distribution function, $g(R, \beta; t|\mathrm{HB})$, which denotes the
distance-angle distribution function of $(R, \beta)$ at time
$t$ for a pair of water molecules
located in the H-bonded region at an initial time $t=0$.
Then, the time dependent ratio of the distance-angle distribution
function is defined as follows,
\begin{equation}
G(R, \beta; t)=\frac{g(R, \beta; t|\mathrm{HB})}{g(R,\beta)}.
\end{equation}
This function characterizes the evolution of the spatial correlations of
the H-bond over time.
Over longer times, the nonequilibrium distribution $g(R, \beta;
t=0|\mathrm{HB})$ initially being inside the H-bond region
finally recovers to that of the equilibrium distribution $g(R, \beta)$ because of the memory loss.
More precisely, we observe $g(R, \beta; t|\mathrm{HB}) \to (N_\mathrm{HB}/N)
g(R, \beta)$ $(t\to \infty)$, where $N$ is the total number
of molecules in the system and
$N_\mathrm{HB}$ denotes the number of averaged accepted H-bonds calculated from integrating
$g(R, \beta)$ over the H-bond region (denoted as region HB),
\begin{equation}
N_\mathrm{HB}=\int\int_{\mathrm{HB}} 2\pi \rho R^2 \sin\beta g(R, \beta)
 dRd\beta.
\end{equation}
Note that $N_\mathrm{HB}$ ranged from from 1.7 ($T=300$ K) to 2.0 ($T=190$ K),
which is not close to 4 since one of the water molecules has
the O atom at the origin and acts only as the H-bond donor, while the
H-bond angle $\beta$ is defined by considering the other water
molecule as the H-bond acceptor.
We also note that the present investigation is analogous to examining
structural relaxation of the selected spectral using the hole-burning technique.~\cite{Laenen:1998dt}

Figure~\ref{fig:diff_oo_fig}(Multimedia view) shows the time series of $G(R, \beta; t)$.
At short time scales, the distribution is firstly elongated towards the angle
direction (increasing $\beta$).
This observation shows that the kinetic pathway of
the H-bond breakage mainly passes through the saddle point on the 2D PMF
profile.
However, as the temperature is decreased, the penetration towards the
distance direction (increasing $R$) becomes more apparent at longer time scales, showing 
transitions that do not pass through the saddle point.
The random, but highly tetrahedral structures lead to frustrations in the
H-bond networks in supercooled water.
It is possible that these frustrations can be relaxed by collective molecular
rearrangements, causing the translational jump
motions.~\cite{Ohmine:1993ij, Stanley:1997uu, Giovambattista:2004ft, Agmon:2012jd}
Such collective rearrangements require larger activation energies.
Furthermore, the 2D PMF using the
distance-angle combinations of two-molecule geometry, $(R, \beta)$, does
not show the 
saddle points consistent with the TST-type behavior.

The translational H-bond breakages are also regarded as the cage effects in
glass-forming liquids, suggesting a transient environment around a 
tagged molecule surrounded by neighboring molecules.~\cite{Doliwa:1998bg, Weeks:2002ch, Schweizer:2004iv,
Shiba:2012hm, PicaCiamarra:2015dz, Pastore:2017hn}
As demonstrated in various studies, there exists a plateau at an
intermediate time regime in the translational mean square
displacement of supercooled water, which is a manifestation of cage
effects.~\cite{Gallo:1996hf, Sciortino:1996hk,
Giovambattista:2004ft, Mazza:2007kr, Kawasaki:2017gw}
In fact, the decay of $c(t)$ at the time regime $\tau_\mathrm{HB}$ is dominated
by the diffusion process.~\cite{Starr:2000cz}
Furthermore, the coupling between 
$\tau_\mathrm{HB}$ and the
translational diffusion constant $D$ was suggested from 
the relationship $D\sim {\tau_\mathrm{HB}}^{-1}$ in
TIP4P/2005 supercooled water.~\cite{Kawasaki:2017gw}

\subsection{Time scales of rotational and translational H-bond breakages}
\label{sec:transition}

For the 2D PMF, it is impracticable to calculate the flux across a
dividing surface as the kinetic pathway is not along a one-dimensional coordinate.
As an alternative, the populations of non H-bond regions adjacent to the H-bonded region were quantified.
For this purpose, we define the rectangular H-bond breakage
regions, as described in Fig.~\ref{fig:2D_PMF}.
The H-bond breakage region due to rotational motions (increasing the angle $\beta$)
is defined as 
$(0.24~\mathrm{nm}, 30^\circ) < (R, \beta) <
(0.34~\mathrm{nm}, 60^\circ)$, where this region is denoted by R.
The H-bond breakage region due to translational motions (increasing
the distance $R$) is also defined as $(0.34~\mathrm{nm},
0^\circ) < (R, \beta) < (0.47~\mathrm{nm}, 30^\circ)$, denoted as region T.
Note that the area of the H-bond breakage region obtained from the integral $2\pi R^2 \sin \beta$
is same for regions R and T.
The population of each region was calculated using:
\begin{align}
N_i(t) = \int\int_{i} 2\pi \rho R^2\sin\beta g(R,
 \beta; t|\mathrm{HB}) dRd\beta,
\end{align}
where $i$ represents the symbol of the region ($i\in$\{HB, R, T\}).
Normalization by 
the number of averaged accepted H-bonds, $N_\mathrm{HB}$, is defined as
$P_i(t)\equiv N_i(t)/N_\mathrm{HB}$.
Note that $P_\mathrm{HB}(t)=N_\mathrm{HB}(t)/H_\mathrm{HB}$ is equivalent to the
H-bond correlation function $c(t)$ from the definition. 
We also note that the sum, $P_\mathrm{HB}(t)+P_\mathrm{R}(t)+P_\mathrm{T}(t)$, is
not a conserved quantity; it begins as unity at $t=0$
($P_\mathrm{HB}(0)=1$ and $P_\mathrm{R}(0)=P_\mathrm{T}(0)=0$) and decays
to zero as the populations inside the regions R and T at time $t$
will spontaneously migrate to other non H-bond states afterwards.

Figure~\ref{fig:population} shows the time evolution of the normalized population,
$P_i(t)(= N_i(t)/N_\mathrm{HB})$.
Both $P_\mathrm{R}(t)$ and $P_\mathrm{T}(t)$ started from zero and
increases with time $t$. 
This regime indicates that the inflow due
to the H-bond breakage exceeds the outflow towards other non H-bond states.
For $P_\mathrm{R}(t)$, we observed additional
peaks at around $\tau_\mathrm{lib}\ls 0.1$ ps, which are independent of
the temperature, attributed to libration motion.
They eventually decay to zero after a peak time, where
the outflow exceeds the inflow for longer time scales.
That is, the maximum peaks are determined by the balance.
Thus, the peak times of $P_\mathrm{R}(t)$ and $P_\mathrm{T}(t)$
(denoted as $\tau_\mathrm{R}$ and $\tau_\mathrm{T}$, respectively)
can be regarded as the characteristic time scales of irreversible
H-bond breakages due to
rotational and translational motions, respectively.
It should be noted that the pathways going from region R(T) to T(R) were
not detected.
In fact, the H-bond
breakages for time regimes at either $\tau_\mathrm{R}$ or
$\tau_\mathrm{T}$ are irreversible for a particular pair of water molecules and
two water molecules hardly reform the same H-bond pair.
We observed that there existed a population of
the translational H-bond breakage, even at higher temperatures.
The peak value of $P_\mathrm{T}(t)$ 
becomes comparable to that of $P_\mathrm{R}(t)$ with decreasing the temperature.
This indicates the increasing number of pathways not going through the saddle
point on the 2D PMF profile, particularly for supercooled water.

\begin{figure}[t]
\centering
\includegraphics[width=0.35\textwidth]{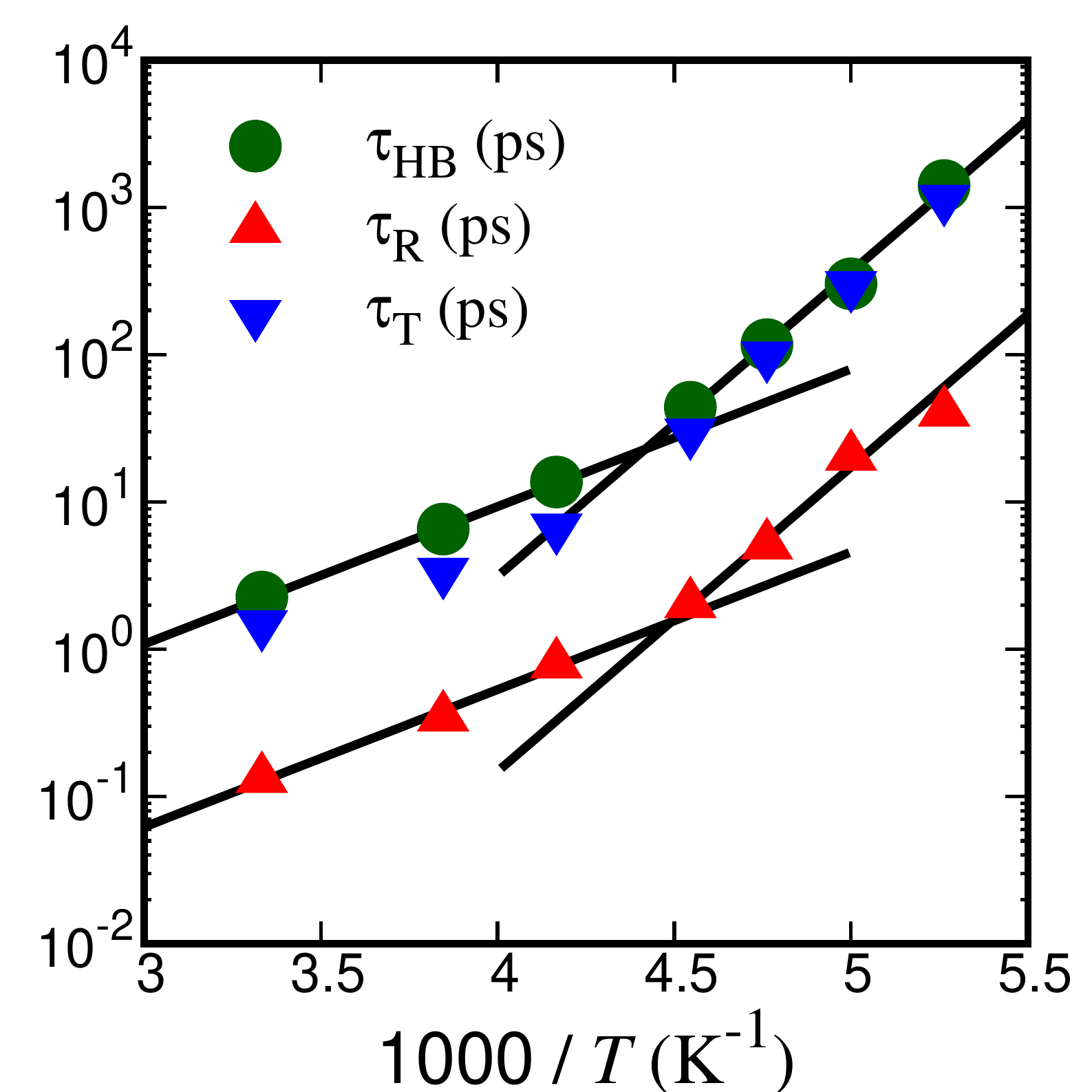}
\caption{
Temperature dependence of the maximum times in $P_\mathrm{R}(t)$ and $P_\mathrm{T}(t)$, as
 denoted by $\tau_\mathrm{R}$ and $\tau_\mathrm{T}$, respectively.
Comparisons with $\tau_\mathrm{HB}$ are also shown.
The straight lines represent Arrhenius behavior $
 \exp(E_\mathrm{A}/k_\mathrm{B}T)$ in the
 high and low temperature ranges with
 activation energies with $E_\mathrm{A} =$ 17.9 and 39.6 kJ/mol, respectively. 
}
\label{fig:population_rate}
\end{figure}

The temperature dependence of $\tau_\mathrm{R}$ and $\tau_\mathrm{T}$ is
plotted in Fig.~\ref{fig:population_rate}.
It can be seen that $\tau_\mathrm{T}$
is comparable with $\tau_\mathrm{HB}$ at all studied temperatures.
In contrast, the time scale of $\tau_\mathrm{R}$ is about one order of
magnitude smaller than $\tau_\mathrm{T}$ although the temperature dependence
of $\tau_\mathrm{R}$ shows similar Arrhenius behaviors.
These observations indicate that the H-bond lifetime $\tau_\mathrm{HB}$ is dominated by the 
H-bond breakages caused by translational motions.
Finally, this observation $\tau_\mathrm{T}\simeq \tau_\mathrm{HB}$
results in the following implication for the temperature dependence of $\kappa$;
\begin{equation}
\frac{1}{\kappa} = \frac{\tau_\mathrm{HB}}{\tau_\mathrm{s}
}\propto \tau_\mathrm{T}
\exp\left(-\frac{C\Delta G^\ddagger }{k_\mathrm{B}T}
    \right).
\end{equation}
This relationship is direct evidence that the temperature dependence of $\kappa$ is
caused by that of the time scale of 
H-bond breakage due to translational motions.

\section{Conclusions}
\label{sec:conclusions}

We analyzed the H-bond breakage dynamics in the TIP4P supercooled water
considering a geometric definition of the H-bond.
We first investigated the temperature dependence of the 2D PMF obtained from the
distance-angle distribution function.
It was found that the position of the saddle point distinguishing
H-bond and non H-bond regions remains unchanged at all the temperatures studied, while
the free-energy barrier of the saddle point, $\Delta G^\ddagger$, gradually increased with
decreasing temperature.
We showed that 
the Arrhenius activation energy of $\tau_\mathrm{HB}$ increases with
decreasing temperature.
In contrast, the TST rate constant $k_\mathrm{TST}$ of the H-bond breakage
approximately followed the Arrhenius temperature dependence.
In addition, 
the TST-type expression, $\exp(-C\Delta
G^\ddagger/k_\mathrm{B}T)$, was obtained, where the correction
factor $C$ was temperature independent.
This suggests a significant decrease in the transmission coefficient
$\kappa$ with decreasing temperature.

To elucidate the molecular mechanism of H-bond breakage,
the kinetic pathways of H-bond breakage were studied.
In particular, the time dependence of the conditional distance-angle
distribution function revealed the pathways that did not go through the
saddle point when the system was deeply supercooled, which was
attributed to 
H-bond breakages due to translational motions.
This translational H-bond breakage is thought to be
associated with the cage-jump
motion, which is commonly observed in various glass-forming
liquids.
In addition, it is proposed that the avalanches of cage-jump motions
trigger collective molecular motions, referred to as dynamic heterogeneities.
Our observations indicated that such collective motions involving
many molecules were not described well by the present 2D PMF using geometric variables
defined by the water dimer.

Furthermore, we quantified the time dependent populations of the
rotational and translational H-bond breakages.
With decreasing temperature, the population of the translational
H-bond breakage became comparable to that of the rotational H-bond
breakage passing through the saddle point.
In particular, the time scale of the translational H-bond breakage
$\tau_\mathrm{T}$ became much longer than that of the rotational H-bond breakage
$\tau_\mathrm{R}$.
The temperature dependence of the H-bond lifetime
$\tau_\mathrm{HB}$ was comparable to that of $\tau_\mathrm{T}$.
This suggests that the H-bond lifetime $\tau_\mathrm{HB}$ is dominated
by the translational motions.

Finally, we note that 
it is important to investigate whether the 2D PMF
associated with H-bonds and their breakages is suitable for supercooled water.
In fact, Sciortino \textit{et al.} demonstrated that
the bifurcated H-bond configuration promotes H-bond breakages with high
mobility~\cite{Sciortino:1991hh, Sciortino:1992eu}, which is apparently
``hidden'' in the 2D PMF drawn for a pair of water molecules; a
bifurcated bond can be explicitly described only beyond the pair level.
Our observations suggested that there exists another possible saddle point
on the 2D PMF profile that reflects the translational H-bond breakage 
relevant with the bifurcated bonds.
Hence, the profile of the PMF might be appropriately transformed, even with the
same reaction coordinates, by using 
information regarding the kinetic pathways from the H-bond to the non H-bond regions.
We are currently undertaking further investigations to clarify this issue.

\begin{acknowledgments}
The authors thank T. Kawasaki, T. Yagasaki,
 T. Joutsuka, and
 Y. Yonetani for helpful discussions.
This work was supported by JSPS KAKENHI Grant Numbers JP16H00829(K.K.), JP18H01188(K.K.), JP15K13550(N.M.), and JP26240045(N.M.).
This work was also supported in part by 
the Post-K Supercomputing Project and the Elements
Strategy Initiative for Catalysts and Batteries from the Ministry of
 Education, Culture, Sports, Science, and Technology.
The numerical calculations were performed at Research Center of Computational
Science, Okazaki Research Facilities, National Institutes of Natural Sciences, Japan.
\end{acknowledgments}

\end{document}